\documentclass[aps,prl,showpacs,amsmath,amssymb,twocolumn]{revtex4}\usepackage{pslatex}

\usepackage{graphicx}
\usepackage{dcolumn}
\usepackage{bm}
\usepackage{natbib}

\begin{document}
\title{Nearly-logarithmic decay of correlations in
glass-forming liquids}

\author{W. G\"otze}
\author{M. Sperl}
\affiliation{Physik-Department, Technische Universit\"at M\"unchen, 
85747 Garching, Germany}

\date{January 23, 2004}
\begin{abstract}
Nearly-logarithmic decay of correlations, which was
observed for several supercooled liquids in 
optical-Kerr-effect experiments [G. Hinze et al. Phys. Rev. Lett. 84, 2437 
(2000), H. Cang et al. Phys. Rev. Lett. 90, 197401 (2003)], is explained 
within the mode-coupling theory for ideal glass transitions
as manifestation of the $\beta$-peak phenomenon. 
A schematic model, which describes the dynamics by only two correlators, one
referring to density fluctuations and the other to the reorientational 
fluctuations of the molecules, yields for strong rotation-translation coupling
response functions in agreement with those measured for benzophenone and Salol 
for the time interval extending from 2~picoseconds to about 20 and 
200~nanoseconds, respectively.
\end{abstract}


\maketitle

Optical-Kerr-effect (OKE) spectroscopy is a powerful technique for the study 
of the dynamics of supercooled liquids \cite{Torre1998}. The 
experiment provides a response function $\chi_A(t)$ for times $t$ exceeding a 
fraction of a picosecond. The function $\chi_A(t)$ is proportional to the 
negative time derivative of a correlator $\phi_A(t)=\langle A^*(t)A\rangle/
\langle |A|^2\rangle$. Here, $\langle\rangle$ denotes canonical averaging, and
the probing variable $A$ is the anisotropic part of the dielectric function. 
The instrumentation was improved recently by application of heterodyne
detection. As a result, it was possible to measure $\chi_A(t)$ for times up to 
500~nanoseconds, i.e., glassy dynamics was documented for the enormous 
time interval starting at the end of the transient and extending over more than
five orders of magnitude \cite{Hinze2000}. It was shown that the evolution of 
the glassy dynamics of Salol upon decreasing the temperature $T$
can be interpreted by the universal formulas 
derived within the mode-coupling theory of ideal glass transitions (MCT)
\cite{Torre1998,Hinze2000}. The fit values for the various parameters have been
found to be consistent with those obtained by other 
light-scattering techniques \cite{Li1992b,Yang1995}. However, the universal 
formulas do not describe the data for all times outside the transient regime. 
Rather, if $T$ decreases, there opens a time interval larger than two orders
of magnitude, which precedes the interval of validity of the universal 
formulas. Here, the response for Salol follows closely a 
$1/t$-law, i.e., the correlator 
exhibits nearly a logarithmic decay: $\phi_A(t)\propto-\ln(t/\tau)$ 
\cite{Hinze2000}. This intriguing feature was observed 
also for some other van-der-Waals liquids, like benzophenone~(BZP). 
This holds with the reservation that the heretofore 
unknown relaxation process can be described more adequately by a power 
law, $\phi_A(t)-f\propto-t^{b'}$, albeit with a rather small exponent 
$b'$ \cite{Cang2003}. Gaining an understanding of the indicated findings
is a challenge to all theories aiming to unlock a comprehensive description 
of liquids. In this Letter, it will be shown that the measured complex 
relaxation scenario \cite{Hinze2000,Cang2003} is a generic, though not 
universal, implication of the standard MCT.

The MCT was proposed as a mathematical model for glassy dynamics 
\cite{Bengtzelius1984}, whose fascinating features are obtained as implication 
of bifurcation points of nonlinear equations of motion derived for the density 
fluctuations. The basic bifurcation is a fold singularity describing a 
transition from ergodic to non-ergodic behavior if the temperature decreases 
through a critical value $T_c$. Using $\epsilon=(T_c-T)/T_c$ as a small 
parameter, the long-time behavior of the correlators can be evaluated 
by asymptotic expansions \cite{Franosch1997}. The leading-order formulas 
establish the universal features of the dynamics. Comparisons of experimental 
results with these formulas \cite{Goetze1999} and tests by molecular-dynamics 
simulation \cite{Goetze1999,Kob2003pre} have shown that MCT is a serious 
candidate for an explanation of glassy dynamics.
Nearly-logarithmic decay of correlations was predicted within MCT for states 
close to bifurcations of the cusp type as can be inferred from 
Ref.~\cite{Goetze2002} and the papers cited there. It was pointed out 
\cite{Hinze2000,Cang2003} that the OKE data might be fitted by the universal 
results for relaxation near cusp bifurcations. However, one has to vary at 
least two control parameters in order to approach such singularity. A fit of 
the OKE data, which only depend on the single control parameter $T$, would 
require implausible fine tuning of the coupling coefficients entering the MCT 
equations. The explanation proposed below is based on the existence of the 
simple liquid-glass-transition singularity. Our results are valid for response 
functions of variables $A$ that couple strongly to density fluctuations so that
a so-called $\beta$-peak can occur in the susceptibility spectra 
\cite{Buchalla1988,Goetze1989c}.

Let us consider an  MCT model describing schematically the density fluctuations
by a single correlator $\phi(t)$. It obeys the Zwanzig-Mori equation of 
motion 
\begin{equation}\label{eq:eom}
\partial_t^2 \phi(t) + \nu \partial_t \phi(t) + \Omega^2 \phi(t) 
+ \Omega^2 \int_0^t m(t-t')\partial_{t'}\phi(t') dt'=0\,,
\end{equation}
where $\Omega$ and $\nu$ are frequencies parameterizing the short-time 
asymptote, $\phi(t)=1-1/2\; (\Omega t)^2 + 1/6\; \nu\Omega^2t^3+{\cal O}(t^4)$.
The kernel $m(t)$, which represents the interactions of the density 
fluctuations, is modeled as
\begin{equation}\label{eq:kernel}
m(t)=v_1\phi(t)+v_2\phi^2(t)\,.
\end{equation}
Here, $v_1\geqslant 0$ and $v_2\geqslant 0$ are the coupling coefficients.
The state of the system is specified by a point $\mathbf{V}=(v_1,v_2)$ in the 
$v_1$-$v_2$ plane. For small $\mathbf{V}$, the model describes liquid states 
where $\phi(t\rightarrow\infty)=0$. For large $\mathbf{V}$, one gets a 
nontrivial long-time limit $f=\phi(t\rightarrow\infty)$, $0<f<1$. Parameter $f$
quantifies the arrest of density fluctuations in the non-ergodic glass state. 
There are lines of liquid-glass transition points $\mathbf{V}^c=(v_1^c,v_2^c)$,
each characterized by some number $\lambda,\,1/2\leqslant\lambda<1$. This 
number determines the critical exponent $a$, $0<a<1/2$, and the von~Schweidler 
exponent $b$, $0<b\leqslant 1$, by the relation 
$\Gamma(1-a)^2/\Gamma(1-2a)=\lambda=\Gamma(1+b)^2/\Gamma(1+2b)$. We consider 
the line, which can be parameterized by
$v_1^c = (2\lambda-1)/\lambda^2\,,\;
v_2^c = 1/\lambda^2\,,\;
0.5\leqslant\lambda<1\,.$
Crossing the line, the long-time limit of $\phi(t)$ jumps from zero to the 
critical value $f^c=1-\lambda$. The distance of the state $\mathbf{V}$ from 
$\mathbf{V}^c$ is specified by the separation parameter
$\sigma = [(v_1-v_1^c)+(v_2-v_2^c)f^c] f^c(1-f^c)\,,$
which is negative for liquid states and positive for glass states. This model 
is the simplest one reproducing all values for the anomalous exponents, which 
can occur in the general theory \cite{Goetze1984}. The parameter $f^c$ appears 
as plateau value of the $\phi(t)$-versus-$\log t$ curves for $\mathbf{V}$ near
$\mathbf{V}^c$ or as relative strength of the $\alpha$-peak of the loss 
spectrum \cite{Franosch1997}.

The probing-variable correlator $\phi_A(t)$ obeys the same general equation of
motion and initial decay as $\phi(t)$, but $\Omega$, $\nu$ and $m(t)$ have to 
be replaced by the corresponding quantities $\Omega_A$, $\nu_A$ and $m_A(t)$. 
The model for the kernel is
\begin{equation}\label{eq:mA} 
m_A(t)=v_A\phi(t)\phi_A(t) \,,
\end{equation}
where $v_A>0$ quantifies the coupling of the probing variable to the density
fluctuations. If $v_A f^c>1$, the long-time limit of $\phi_A(t)$ jumps from 
zero to the critical value $f_A^c = 1-1/(v_A f^c)$ as
$\mathbf{V}$ crosses the transition line. Originally, this model 
was motivated for the density fluctuations of a tagged particle 
\cite{Sjoegren1986}. Also, the MCT equations for reorientational 
dynamics of a linear molecule suggest an expression like Eq.~(\ref{eq:mA})
\cite{Chong2002b}. Within the microscopic theory, the coupling coefficients 
depend on the equilibrium structure functions, which depend smoothly on $T$. 
Therefore, $v_1$, $v_2$, and $v_A$ are smooth functions of the temperature. The
specified model was applied repeatedly for the description of experiments, 
as can be inferred from Ref.~\cite{Goetze2000b} and the papers cited there.

The long-time decay of the correlators at the transition point is given by 
$[\phi(t)-f^c]/h =[\phi_A(t)-f_A^c]/h_A = (t_0/t)^a+{\cal O}(t^{-2a})$. The 
time $t_0$ depends on the transient dynamics. It is determined by matching the 
numerical solution of $\phi(t)$ for large times to the asymptotic 
formula. The amplitudes are $h=\lambda\,,\;h_A=\lambda/(v_A f^{c\,2})$. In the 
limit of small $|\sigma|$, there appears a large time interval where 
$|\phi(t)-f^c|$ is small. Here, one gets in leading-order $[\phi(t)-f^c]/h = 
[\phi_A(t)-f_A^c]/h_A = c_\sigma g(t/t_\sigma)$. This is a scaling law with a 
correlation scale $c_\sigma = \sqrt{|\sigma|}$ and a time scale $t_\sigma= 
t_0/|\sigma|^{1/2a}$. The function $g(\hat{t})$ is determined solely by 
$\lambda$. It exhibits the critical power law for small rescaled times 
$\hat{t}$, $g(\hat{t}\ll1)=1/\hat{t}^a$, and von~Schweidler's power law for 
large rescaled times in the liquid, $g(\hat{t}\gg1) = -B\hat{t}^b$ 
\cite{Franosch1997}. The scaling law implies one for the response functions
\begin{equation}\label{eq:chi}
\chi(t)/h=\chi_A(t)/h_A = s_\sigma k(t/t_\sigma)\,,
\end{equation}
where $s_\sigma=c_\sigma/t_\sigma$. 
The master function $k(\hat{t}) = -\partial \hat{g}/\partial\hat{t}$ 
interpolates between the critical power law,
$k(\hat{t}\ll1) = -a /\hat{t}^{(1+a)}\,,$
and, in the liquid state, the von~Schweidler power law,
$k(\hat{t}\gg1) = -b B/\hat{t}^{(1-b)}\,.$

The VH-light-scattering spectra of Salol \cite{Li1992b} or orthoterphenyl 
\cite{Cummins1997} show that $f_A^c$ is about $0.9$. This means there is strong
arrest of the reorientational motion at the ideal glass-transition points in 
these van-der-Waals systems. Integrating $\chi_A(t)$ over time, one can 
determine $\phi_A(t)$ and read off the plateau. We find $f_A^c=0.90\pm0.05$ 
from the $T=260\text{K}$-data for BZP \cite{Cang2003} and $f_A^c=0.93\pm0.03$ 
from those for Salol at $T=257\text{K}$ \cite{Hinze2000}, corroborating the 
preceding conclusion. Therefore, we focus on $\chi_A(t)$ for large 
translation-rotation coupling $v_A$.

\begin{figure}[htb]
\includegraphics[width=0.9\columnwidth]{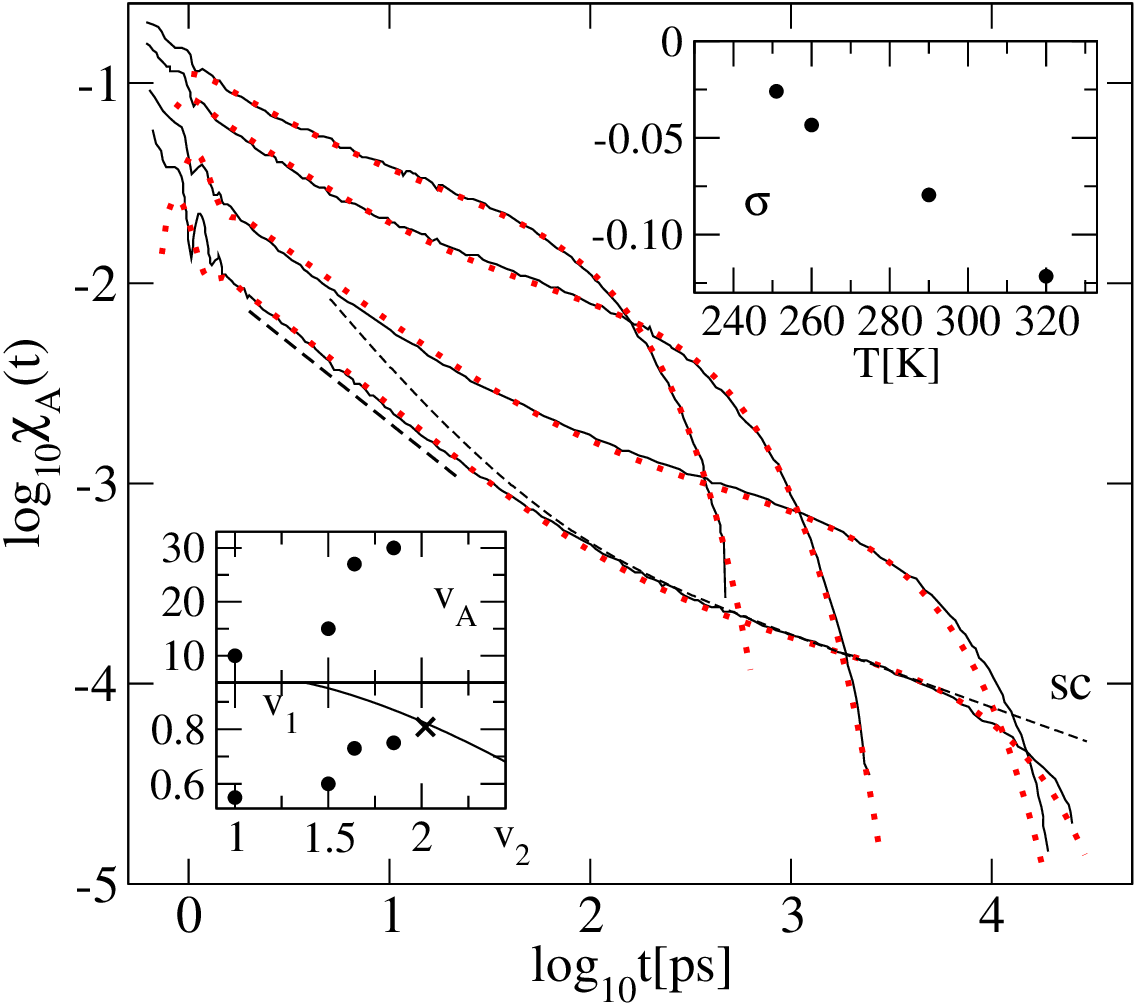}
\caption{\label{fig:BZP}OKE response $\chi_A(t)$
measured for BZP for temperatures $T/\text{K}=251,\,260,\,290,\,320$ (full
lines from bottom to top) \cite{Cang2003}. The dashed straight line has slope 
$x=0.80$. The dotted lines exhibit results $N\cdot\chi_A(t)$, with 
$\chi_A(t)$ calculated for the MCT model defined by 
Eqs.~(\ref{eq:eom}--\ref{eq:mA}) with $\Omega=\Omega_A=0.2\nu=0.2\nu_A$.  The 
factors $N$ and $\Omega$ are chosen (from bottom to top) as 3.5,
6, 9.5, 8, and 1.67, 1.43, 2.0, 2.0 THz, respectively. The left inset
exhibits the variation of the coupling constants with changes of the 
temperature, and the right inset the one of the separation parameter $\sigma$.
The dashed line marked {sc} shows a scaling-law result, 
Eq.~(\ref{eq:chi}), calculated for $\lambda=0.70$. The line in the lower left 
inset exhibits transition points $\mathbf{V}^c$; the point
$\sigma=-0.003$ for $\lambda=0.70$, is marked by a cross.
}
\end{figure}

Figure~\ref{fig:BZP} reproduces the OKE response functions for BZP 
\cite{Cang2003}. A dashed straight line of slope $x=0.80$ shows that the data 
for $T=251~\text{K}$ exhibit power-law decay $\chi(t)\propto1/t^x$ for 
2~ps$\leqslant t\leqslant20$~ps. There is a von~Schweidler-law-like variation: 
$\phi(t)-\text{const.}\propto-t^{b'},\,b'=1-x=0.20$. In the double-logarithmic 
representation, the scaling-law result, Eq.~(\ref{eq:chi}), appears as 
interpolation between a straight line of slope $(1+a)>1$, and one of slope 
$(1-b)<1$. Changing the scales $s_\sigma h_A$ and $t_\sigma$ is equivalent to a
translation of the curve. The dashed line marked sc shows an example calculated
for $\lambda=0.70$ ($a=0.33,\,b=0.64$) shifted to match the data for 
$T=251~\text{K}$. It provides a proper description of the observations for 
$60~\text{ps}<t<6~\text{ns}$. However, since $1+a>1>x$, the scaling law cannot 
account for the data for $t<60$~ps. This conclusion \cite{Cang2003} is not 
altered by choosing $\lambda$ or $t_\sigma$, $s_\sigma$ differently. 
The dots are the 
theoretical results for $\chi_A(t)$ scaled by amplitude factors $N$ and scales 
$\Omega$, which vary somewhat with $T$. Notice that the calculated functions 
exhibit transient oscillations for times below and up to 2~ps, as do the 
measured curves. This shows that the parameters $\Omega,\, \nu,\, \Omega_A,\, 
\nu_A$ are chosen reasonably. The strong variation of $\chi_A(t)$ with changes 
of $T$ is due to the changes of the coupling coefficients as documented in the 
left inset. The state $\mathbf{V}=(v_1,v_2)$ shifts towards the transition line
if $T$ decreases. An extrapolation to a transition point for $\lambda=0.70$, 
which is near the cross, is consistent with the fit values for $v_1$ and $v_2$.
Parameter $\sigma$ exhibits an almost linear temperature dependence, and 
extrapolation to $\sigma=0$ suggests $T_c\approx 235~\text{K}$. Uncertainty 
estimates for $\lambda$ and $T_c$ cannot be given yet, since we did not study 
in detail the possibility for data fits by other choices for $v_1$, $v_2$, and 
$v_A$.

\begin{figure}[htb]
\includegraphics[width=0.9\columnwidth]{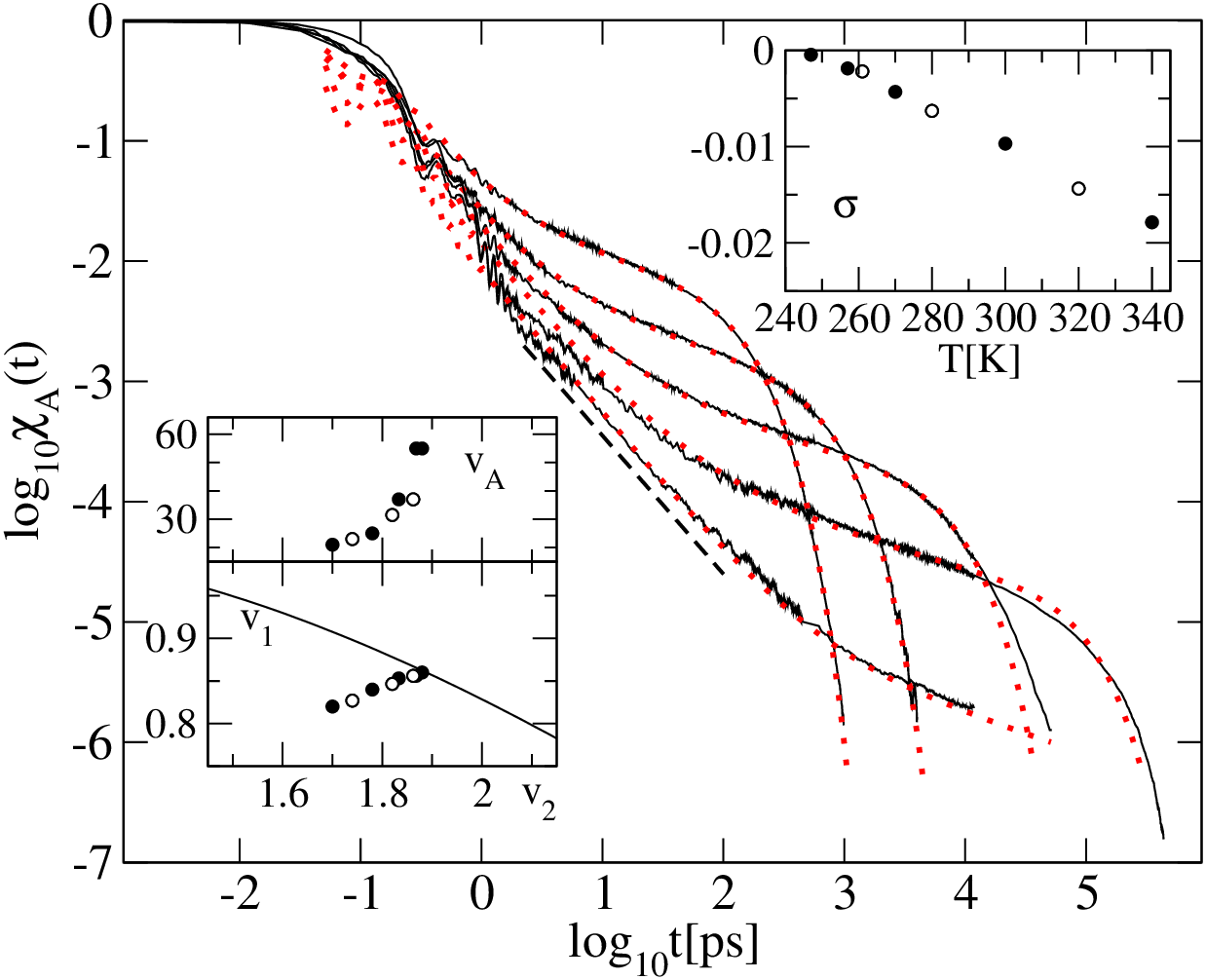}
\caption{\label{fig:salol}Analog results as the ones in Fig.~\ref{fig:BZP}
for data measured for Salol for $T/\text{K}=247,\, 257,\,270,\, 300,\, 340$ 
\cite{Hinze2000}. The dashed straight line has slope x=1.15. The MCT results 
are calculated for $7.9~\text{THz} = \Omega = 2\Omega_A = 10\nu_A,\,\nu=0$, and
factors $N=$ 25, 37.3, 27.8, 17.9, 19, respectively. The open circles in the 
insets are the fit values for the other measured temperatures \cite{Hinze2000};
the corresponding response functions are not shown to avoid overloading.
}
\end{figure}
Figure~\ref{fig:salol} shows the OKE response for Salol for the 
highest and lowest temperature measured and three  
temperatures in between \cite{Hinze2000}. The data are normalized to unity 
for $t=0.01$~ps. A dashed straight line of slope $x=1.15$ shows that the 
data for $T=247~\text{K}$ can be described by $\chi(t)\propto1/t^x$ for the 
interval 2~ps $<t<$ 100~ps. Equivalently, the correlator shows a decay 
similar to the critical one, $\phi(t)\propto 1/t^{a'}$ with $a'=x-1=0.15$. 
For the $T=257$~K data, a similar behavior is found with $x\approx 0$, 
i.e., $\phi(t)$ exhibits logarithmic decay \cite{Hinze2000}. 
The MCT fits have been evaluated 
for temperature-independent parameters $\Omega,\, \Omega_A,\, \nu,\, 
\nu_A$. The normalization constants $N$ exhibit some $T$ dependence. The 
path $\mathbf{V}=(v_1,v_2)$ extrapolates to a critical point for $\lambda=0.73$
($a=0.31,\,b=0.59$). The $\sigma$-versus-$T$ results suggest $T_c\approx 
245~\text{K}$. The numbers for $\lambda$ and $T_c$ are consistent with those 
obtained previously \cite{Torre1998,Hinze2000,Li1992b,Yang1995}. For the 
identification of the ideal transition point it would be helpful to have data 
available also for $T$ below the estimated $T_c$. The variation
of $v_A$ closely follows an Arrhenius law: $v_A\propto\exp(T_0/T)$, where the 
fit value $T_0=1000~\text{K}$ is not unreasonable for a van-der-Waals liquid.

\begin{figure}[htb]
\includegraphics[width=0.9\columnwidth]{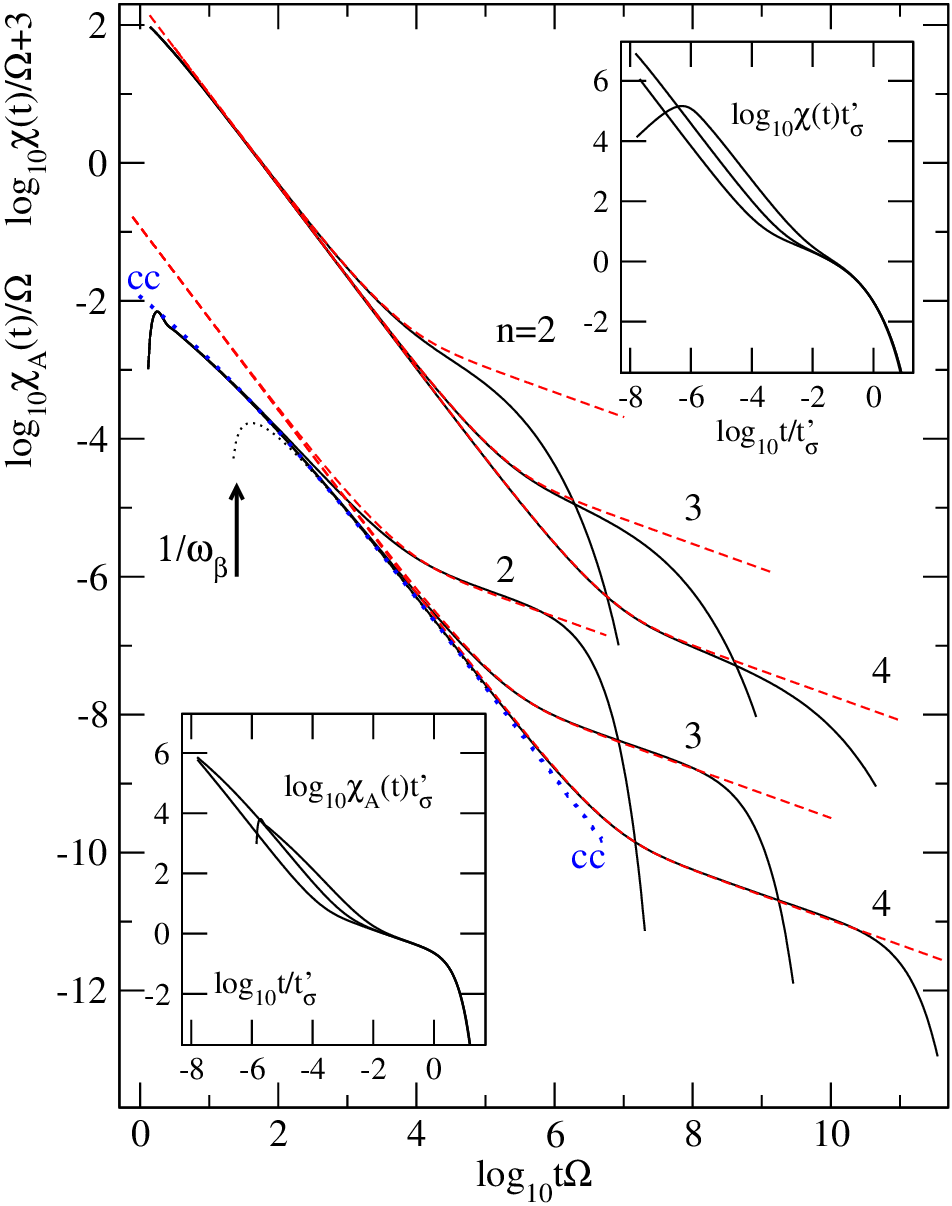}
\caption{\label{fig:scaling}Response functions for the model defined by 
Eqs.~(\ref{eq:eom}--\ref{eq:mA}) calculated with $\Omega=\Omega_A=\nu/5 = 
\nu_A/5$ and $v_A=20$ for three states specified by $\lambda=0.70$, $(v_1,v_2)=
 (v_1^c,v_2^c) (1+\varepsilon),\, \varepsilon = -10^{-n},\, n=2,\,3,\,4$, 
yielding $\sigma=3\varepsilon/10$. The results for $\chi(t)$ are shifted 
upwards by 3~decades. The dashed lines are the scaling-law results, 
Eq.~(\ref{eq:chi}), where $t_0=0.755$. The dotted line shows the leading 
correction to the scaling-law for short times. The dotted line marked cc 
exhibits the Cole-Cole response function $\chi_A^\text{cc}(t)$. The insets 
demonstrate the superposition principle for the long-time dynamics.
}
\end{figure}
Figure~\ref{fig:scaling} shows the two response functions, $\chi(t) = 
-\partial\phi(t)/\partial t$ and  $\chi_A(t) = -\partial\phi_A(t)/\partial t$,
for three liquid states near the transition point with $\lambda=0.70$ and the
corresponding scaling-law results. There is a second scaling law, the 
superposition principle for the $\alpha$~process, dealing with the dynamics for
times within the von~Schweidler-law regime and longer: $\chi(t) =
\tilde{\chi}(t/t'_\sigma)/t'_\sigma, \,\chi_A(t) =
\tilde{\chi}_A(t/t'_\sigma)/t'_\sigma$,
$t'_\sigma=t_0/|\sigma|^\gamma, \, \gamma=1/(2a)+1/(2b)$. The insets 
demonstrate this result. The solution at $\mathbf{V}^c$ including the leading 
correction reads $[\phi_A(t)-f_A^c] = h_A (t_0/t)^a [1+K_A(t_0/t)^a]$ and a 
corresponding formula with index $A$ dropped. The number $K_A=-1.5$ is 
evaluated from the coupling constants $v_1$, $v_2$, $v_A$; it is negative and 
large since $f_A^c$ is high \cite{Franosch1997}. The result is shown as dotted 
line for $\chi_A$. The corresponding correction for $\chi(t)$, $K=0.020$, is so
small that it cannot be made visible in Fig.~\ref{fig:scaling}. This 
exemplifies that the range of validity of the universal formulas can be 
quite different for different functions. The scaling-law formulas 
explain quantitatively the evolution of the glassy dynamics as exhibited 
by $\chi(t)$ and this for the complete regime 
outside the transient, $\Omega t>2$. The state $n=2$ is marked by a cross 
in the left inset of Fig.~\ref{fig:BZP}. The states of interest for the 
data interpretation have a much larger separation from the critical point. One 
finds here, as in previous work \cite{Goetze2000b,Franosch1997}, 
that the scaling-law description can explain all the qualitative features of 
$\chi(t)$, but that a quantitative description does not work for 
$|\sigma|> 0.05$. The same conclusions are valid for $\chi_A(t)$, 
albeit only for times with $t\Omega\gtrapprox 3000$. The size of the correction
term $K_A$ is so large, that there appears a three-decade time interval 
outside the transient where the universal formulas cannot account for the 
MCT solution. Within the interval $2\leqslant\log_{10}(\Omega t) 
\leqslant 3.5$, the leading-order-correction formula describes the 
results. But for $0.5\leqslant\log_{10}(\Omega t)\leqslant 2$, even this 
formula is insufficient. Within the interval $0.5\leqslant\log_{10} 
(\Omega t)\leqslant 3.5$, nearly-logarithmic decay is exhibited by 
$\phi_A(t)$. 

The equation of motion for $\phi_A(t)$ is equivalent to $\chi_A(\omega) = 
-\Omega_A^2/[\omega(\omega+i\nu_A)-\Omega_A^2+\Omega_A\omega\ m_A(\omega)]$, 
where $\chi_A(\omega)$ and $m_A(\omega)$ are the Fourier transforms of 
$\chi_A(t)$ and $m_A(t)$, respectively, for frequency $\omega$. 
For low-frequency phenomena,
$|\omega(\omega+i\nu_A)|\ll\Omega_A^2$, this simplifies to $\chi_A(\omega) = 
1/[1-\omega\ m_A(\omega)]$. For $t/t_0\gg 1$ and $t/t_\sigma\ll 1$, one gets up
to next-to-leading order: $m_A(t) = f_m +h_m (t_0/t)^a[1+K_m(t_0/t)^a]$, where 
$f_m=v_Af^cf_A^c$, $h_m=v_A(f^ch_A+f_A^ch)$, 
$K_mh_m=v_A[f^ch_AK_A+f_A^chK+hh_A]$. There is a trend that the negative
contribution of the first term in the bracket is canceled by the positive one 
due to the last term. Hence, because the correction term $K_A$ for $\phi_A(t)$
is large, the correction term $K_m=0.16$ for $m_A(t)$ is small. 
Thus, the critical law is a good approximation for $m_A(t)$. This leads
to the Cole-Cole formula for the $\beta$-process: $\chi_A^\text{cc}(\omega) = 
\chi_0^\text{cc}/[1+(-i\omega/\omega_\beta)^a]$, $\chi_0^\text{cc} = 
1/[1+f^cf_A^cv_A]$, $\omega_\beta = [(1+f_m)/h_m\Gamma(1-a)]^{1/a}/t_0$. The 
limit $\omega/\omega_\beta\ll 1$ reproduces the universal critical decay: 
$\chi(t)\propto 1/t^{1+a}$. The limit $\omega/\omega_\beta\gg 1$ leads to a 
von~Schweidler-law-like decay: $\chi(t)\propto 1/t^{1-a}$. It depends on 
the relative size of $t_0^{-1}$, $\omega_\beta$ and $t_\sigma^{-1}$, which part
of the $\beta$-peak spectrum dominates the nearly-logarithmic decay, 
$\chi(t)\propto1/t^x$, $1-a\leqslant x\leqslant 1+a$. The dotted line in 
Fig.~\ref{fig:scaling} marked cc exhibits $\chi_A^\text{cc}(t)$ for our model.
The preceding reasoning for using the leading asymptotic expansion for $m_A(t)$
rather than for $\phi_A(t)$ is valid also for the microscopic version of MCT. 
The schematic model used here is merely the simplest example illustrating 
our derivation of the $\beta$-peak phenomenon.

We have shown that the evolution of glassy dynamics as
measured by OKE spectroscopy for two liquids can be
described reasonably by the solutions of a standard
schematic MCT model, and this for time intervals extending up to five
orders of magnitude. The long-time part of the response functions can be
explained qualitatively by the known scaling laws reflecting leading-order
asymptotic solutions of the MCT equations. The leading-order correction
formulas explain that for strong probing-variable-environment coupling
there appears a large time interval adjacent to the one for the transient
motion where the scaling laws cannot explain the data. The nearly-logarithmic 
decay within this interval is explained as $\beta$-peak dynamics and is fitted 
perfectly by the solutions of the schematic model. Our theory implies the 
prediction of nearly-logarithmic decay of reorientational correlations for
liquids of molecules with large elongation. Molecular-dynamics simulations can 
test this and provide the structure factors for a microscopic calculation 
within MCT. Such work could substantiate or falsify the preceding explanation.
It remains to be explained why the $\beta$-peak phenomenon has not been
detected in earlier studies.

We thank M.D.~Fayer and G.~Hinze for providing their data sets. We thank
them as well as H.Z.~Cummins, V.N.~Novikov, R.~Schilling and Th.~Voigtmann for 
stimulating discussions. Our work was supported by the DFG Grant No. 
Go154/13-2.

\bibliographystyle{apsrev}

\begin{thebibliography}{17}
\expandafter\ifx\csname natexlab\endcsname\relax\def\natexlab#1{#1}\fi
\expandafter\ifx\csname bibnamefont\endcsname\relax
  \def\bibnamefont#1{#1}\fi
\expandafter\ifx\csname bibfnamefont\endcsname\relax
  \def\bibfnamefont#1{#1}\fi
\expandafter\ifx\csname citenamefont\endcsname\relax
  \def\citenamefont#1{#1}\fi
\expandafter\ifx\csname url\endcsname\relax
  \def\url#1{\texttt{#1}}\fi
\expandafter\ifx\csname urlprefix\endcsname\relax\def\urlprefix{URL }\fi
\providecommand{\bibinfo}[2]{#2}
\providecommand{\eprint}[2][]{\url{#2}}

\bibitem[{\citenamefont{Torre et~al.}(1998)\citenamefont{Torre, Bartolini, and
  Pick}}]{Torre1998}
\bibinfo{author}{\bibfnamefont{R.}~\bibnamefont{Torre}},
  \bibinfo{author}{\bibfnamefont{P.}~\bibnamefont{Bartolini}},
  \bibnamefont{and} \bibinfo{author}{\bibfnamefont{R.~M.} \bibnamefont{Pick}},
  \bibinfo{journal}{Phys.~Rev.~E} \textbf{\bibinfo{volume}{57}},
  \bibinfo{pages}{1912} (\bibinfo{year}{1998}).

\bibitem[{\citenamefont{Hinze et~al.}(2000)\citenamefont{Hinze, Brace, Gottke,
  and Fayer}}]{Hinze2000}
\bibinfo{author}{\bibfnamefont{G.}~\bibnamefont{Hinze}},
  \bibinfo{author}{\bibfnamefont{D.~D.} \bibnamefont{Brace}},
  \bibinfo{author}{\bibfnamefont{S.~D.} \bibnamefont{Gottke}},
  \bibnamefont{and} \bibinfo{author}{\bibfnamefont{M.~D.} \bibnamefont{Fayer}},
  \bibinfo{journal}{Phys.~Rev.~Lett.} \textbf{\bibinfo{volume}{84}},
  \bibinfo{pages}{2437} (\bibinfo{year}{2000}), 4783(E);
  \bibinfo{journal}{J.~Chem.~Phys.} \textbf{\bibinfo{volume}{113}},
  \bibinfo{pages}{3723} (\bibinfo{year}{2000}{\natexlab{b}}).

\bibitem[{\citenamefont{Li et~al.}(1992)\citenamefont{Li, Du, Sakai, and
  Cummins}}]{Li1992b}
\bibinfo{author}{\bibfnamefont{G.}~\bibnamefont{Li}},
  \bibinfo{author}{\bibfnamefont{W.~M.} \bibnamefont{Du}},
  \bibinfo{author}{\bibfnamefont{A.}~\bibnamefont{Sakai}}, \bibnamefont{and}
  \bibinfo{author}{\bibfnamefont{H.~Z.} \bibnamefont{Cummins}},
  \bibinfo{journal}{Phys.~Rev.~A} \textbf{\bibinfo{volume}{46}},
  \bibinfo{pages}{3343} (\bibinfo{year}{1992}).

\bibitem[{\citenamefont{Yang and Nelson}(1995)}]{Yang1995}
\bibinfo{author}{\bibfnamefont{Y.}~\bibnamefont{Yang}} \bibnamefont{and}
  \bibinfo{author}{\bibfnamefont{K.~A.} \bibnamefont{Nelson}},
  \bibinfo{journal}{Phys.~Rev.~Lett.} \textbf{\bibinfo{volume}{74}},
  \bibinfo{pages}{4883} (\bibinfo{year}{1995}).

\bibitem[{\citenamefont{Cang et~al.}(2003)\citenamefont{Cang, Novikov, and
  Fayer}}]{Cang2003}
\bibinfo{author}{\bibfnamefont{H.}~\bibnamefont{Cang}},
  \bibinfo{author}{\bibfnamefont{V.~N.} \bibnamefont{Novikov}},
  \bibnamefont{and} \bibinfo{author}{\bibfnamefont{M.~D.} \bibnamefont{Fayer}},
  \bibinfo{journal}{Phys.~Rev.~Lett.} \textbf{\bibinfo{volume}{90}},
  \bibinfo{pages}{197401} (\bibinfo{year}{2003});
  \bibinfo{journal}{J.~Chem.~Phys.} \textbf{\bibinfo{volume}{118}},
  \bibinfo{pages}{2800} (\bibinfo{year}{2003}{\natexlab{b}}).

\bibitem[{\citenamefont{Bengtzelius et~al.}(1984)\citenamefont{Bengtzelius,
  G\"otze, and Sj\"olander}}]{Bengtzelius1984}
\bibinfo{author}{\bibfnamefont{U.}~\bibnamefont{Bengtzelius}},
  \bibinfo{author}{\bibfnamefont{W.}~\bibnamefont{G\"otze}}, \bibnamefont{and}
  \bibinfo{author}{\bibfnamefont{A.}~\bibnamefont{Sj\"olander}},
  \bibinfo{journal}{J.~Phys.~C} \textbf{\bibinfo{volume}{17}},
  \bibinfo{pages}{5915} (\bibinfo{year}{1984}).

\bibitem[{\citenamefont{Franosch et~al.}(1997)\citenamefont{Franosch, Fuchs,
  G\"otze, Mayr, and Singh}}]{Franosch1997}
\bibinfo{author}{\bibfnamefont{T.}~\bibnamefont{Franosch}},
  \bibinfo{author}{\bibfnamefont{M.}~\bibnamefont{Fuchs}},
  \bibinfo{author}{\bibfnamefont{W.}~\bibnamefont{G\"otze}},
  \bibinfo{author}{\bibfnamefont{M.~R.} \bibnamefont{Mayr}}, \bibnamefont{and}
  \bibinfo{author}{\bibfnamefont{A.~P.} \bibnamefont{Singh}},
  \bibinfo{journal}{Phys.~Rev.~E} \textbf{\bibinfo{volume}{55}},
  \bibinfo{pages}{7153} (\bibinfo{year}{1997}).

\bibitem[{\citenamefont{G\"otze}(1999)}]{Goetze1999}
\bibinfo{author}{\bibfnamefont{W.}~\bibnamefont{G\"otze}},
  \bibinfo{journal}{J.~Phys.: Condens.~Matter} \textbf{\bibinfo{volume}{11}},
  \bibinfo{pages}{A1} (\bibinfo{year}{1999}).

\bibitem[{\citenamefont{Kob}(2003)}]{Kob2003pre}
\bibinfo{author}{\bibfnamefont{W.}~\bibnamefont{Kob}}, in
  \emph{\bibinfo{booktitle}{Slow Relaxations and Nonequilibrium Dynamics in
  Condensed Matter}}, edited by \bibinfo{editor}{\bibfnamefont{J.-L.}
  \bibnamefont{Barrat}},
  \bibinfo{editor}{\bibfnamefont{M.}~\bibnamefont{Feigelman}},
  \bibinfo{editor}{\bibfnamefont{J.}~\bibnamefont{Kurchan}}, \bibnamefont{and}
  \bibinfo{editor}{\bibfnamefont{J.}~\bibnamefont{Dalibard}}
  (\bibinfo{publisher}{Springer}, \bibinfo{address}{Berlin},
  \bibinfo{year}{2003}), p. \bibinfo{pages}{199}.

\bibitem[{\citenamefont{G\"otze and Sperl}(2002)}]{Goetze2002}
\bibinfo{author}{\bibfnamefont{W.}~\bibnamefont{G\"otze}} \bibnamefont{and}
  \bibinfo{author}{\bibfnamefont{M.}~\bibnamefont{Sperl}},
  \bibinfo{journal}{Phys.~Rev.~E} \textbf{\bibinfo{volume}{66}},
  \bibinfo{pages}{011405} (\bibinfo{year}{2002}).

\bibitem[{\citenamefont{Buchalla et~al.}(1988)\citenamefont{Buchalla, Dersch,
  G\"otze, and Sj\"ogren}}]{Buchalla1988}
\bibinfo{author}{\bibfnamefont{G.}~\bibnamefont{Buchalla}},
  \bibinfo{author}{\bibfnamefont{U.}~\bibnamefont{Dersch}},
  \bibinfo{author}{\bibfnamefont{W.}~\bibnamefont{G\"otze}}, \bibnamefont{and}
  \bibinfo{author}{\bibfnamefont{L.}~\bibnamefont{Sj\"ogren}},
  \bibinfo{journal}{J.~Phys.~C} \textbf{\bibinfo{volume}{21}},
  \bibinfo{pages}{4239} (\bibinfo{year}{1988}).

\bibitem[{\citenamefont{G\"otze and Sj\"ogren}(1989)}]{Goetze1989c}
\bibinfo{author}{\bibfnamefont{W.}~\bibnamefont{G\"otze}} \bibnamefont{and}
  \bibinfo{author}{\bibfnamefont{L.}~\bibnamefont{Sj\"ogren}},
  \bibinfo{journal}{J.~Phys.: Condens.~Matter} \textbf{\bibinfo{volume}{1}},
  \bibinfo{pages}{4183} (\bibinfo{year}{1989}).

\bibitem[{\citenamefont{G\"otze}(1984)}]{Goetze1984}
\bibinfo{author}{\bibfnamefont{W.}~\bibnamefont{G\"otze}},
  \bibinfo{journal}{Z.~Phys.~B} \textbf{\bibinfo{volume}{56}},
  \bibinfo{pages}{139} (\bibinfo{year}{1984}).

\bibitem[{\citenamefont{Sj\"ogren}(1986)}]{Sjoegren1986}
\bibinfo{author}{\bibfnamefont{L.}~\bibnamefont{Sj\"ogren}},
  \bibinfo{journal}{Phys.~Rev.~A} \textbf{\bibinfo{volume}{33}},
  \bibinfo{pages}{1254} (\bibinfo{year}{1986}).

\bibitem[{\citenamefont{Chong and G\"otze}(2002)}]{Chong2002b}
\bibinfo{author}{\bibfnamefont{S.-H.}~\bibnamefont{Chong}} \bibnamefont{and}
\bibinfo{author}{\bibfnamefont{W.}~\bibnamefont{G\"otze}},
  \bibinfo{journal}{Phys.~Rev.~E} \textbf{\bibinfo{volume}{65}},
  \bibinfo{pages}{051201} (\bibinfo{year}{2002}).

\bibitem[{\citenamefont{G\"otze and Voigtmann}(2000)}]{Goetze2000b}
\bibinfo{author}{\bibfnamefont{W.}~\bibnamefont{G\"otze}} \bibnamefont{and}
  \bibinfo{author}{\bibfnamefont{T.}~\bibnamefont{Voigtmann}},
  \bibinfo{journal}{Phys.~Rev.~E} \textbf{\bibinfo{volume}{61}},
  \bibinfo{pages}{4133} (\bibinfo{year}{2000}).

\bibitem[{\citenamefont{Cummins et~al.}(1997)\citenamefont{Cummins, Li, Du,
  Hwang, and Shen}}]{Cummins1997}
\bibinfo{author}{\bibfnamefont{H.~Z.} \bibnamefont{Cummins}},
  \bibinfo{author}{\bibfnamefont{G.}~\bibnamefont{Li}},
  \bibinfo{author}{\bibfnamefont{W.}~\bibnamefont{Du}},
  \bibinfo{author}{\bibfnamefont{Y.~H.} \bibnamefont{Hwang}}, \bibnamefont{and}
  \bibinfo{author}{\bibfnamefont{G.~Q.} \bibnamefont{Shen}},
  \bibinfo{journal}{Prog.~Theor.~Phys. Suppl.} \textbf{\bibinfo{volume}{126}},
  \bibinfo{pages}{21} (\bibinfo{year}{1997}).

\end{thebibliography}

\end{document}